\documentstyle[12pt,aasms4]{article}

\begin{document}

\title{The Montreal Blue Galaxy Survey III.  
\\ Third list of UV-bright candidates\altaffilmark{1}}

\author{Roger Coziol\altaffilmark{2}}
\affil{Instituto Nacional de Pesquisas Espaciais (INPE), Divis\~ao de Astrof\'{\i}sica\\
Caixa Postal 515, 12201/970 Sao Jose dos Campos, SP, Brazil.} 
\authoremail{coziol@das.inpe.br}
\author{Serge Demers\altaffilmark{2} and R\'emi Barn\'eoud}
\affil{D\'epartement de Physique, Observatoire du Mont M\'egantic\\
Universit\'e de Montr\'eal, Montr\'eal, Qu\'ebec, H3C 3J7 Canada}
\authoremail{demers@astro.umontreal.ca}
\and
\author{Miriam Pe\~na}
\affil{Instituto de Astronomia, Universidad Nacional Aut\'onoma de M\'exico\\
Apdo. Postal 70-264, 04510 M\'exico D. F., M\'exico}
\authoremail{miriam@astroscu.unam.mx}


\altaffiltext{1}{ Partially based on observations obtained at 
Observatorio Astr\'onomico Nacional, San Pedro M\'artir, B. C., M\'exico.}
\altaffiltext{2}{Visiting Astronomer, Cerro Tololo Inter-American Observatory. 
CTIO is operated by AURA, Inc.\ under contract to the National Science
Foundation.}

\begin{abstract}

We present and discuss the latest addition 
of the Montreal Blue Galaxy (MBG) survey.
Inspection of 59 Curtis Schmidt plates resulted in the identification 
of 135 new UV-bright galaxies with B$ < 15.5$. 
This brings the total number of MBGs to 469. 
New results of the $V/V_m$ test show that our survey 
is complete to B = 14.7.    

From our most recent spectroscopic follow--up, we confirm the discovery 
of one new Seyfert 1 galaxy and possibly one new Seyfert 2 galaxy.
We confirm also the bias of the MBG survey towards the low-excitation and metal 
rich Starburst Nucleus Galaxies (SBNGs). 
The spectral characteristics of the MBGs are similar to those 
of the infrared luminous IRAS galaxies.  
As a common characteristic, they show a mean ratio Log([NII]/H$\alpha$) 
in excess of 0.2 dex as compared to normal disk HII regions. 
In general, the MBGs have lower  
far--infrared luminosities (L$_{IR} < 10^{11}$ L$_{\odot}$) 
and are nearer (z $< 0.05$) than the luminous IRAS galaxies. 

The distribution of the morphologies 
of the MBGs indicates a high number of early-type spirals (Sb and earlier). 
Nearly half of these galaxies also possess a bar. 
In our sample, the fraction of galaxies with bars 
depends on the morphology and increases towards the late-type
spirals. However, if we consider only isolated galaxies, 
the late-type spirals show a clear tendency to be barred.  
Signs of a recent interaction 
with neighbor galaxies are obvious only in 24\% of our candidates. 
Although this number is only a lower limit, 
it is nevertheless sufficiently low to suggest that 
in a majority of massive galaxies
the burst of star formation do not depends solely on dynamical processes. 

\end{abstract}

\section{INTRODUCTION}

This article is the third of a series of papers presenting and discussing
the results of the MBG survey.
The aim of our project is to increase substantially the number of known 
UV--bright galaxies in the south galactic pole area.
In parallel with our main effort, spectroscopic and imaging programs have 
also been implemented to determine the nature of our candidates and 
identify the origin of their activity. 
From spectroscopy, we determined that 95\% of the MBGs are starburst galaxies
(Coziol\markcite{c8} {\em et al.} 1993, hereafter Paper~I). 
Comparison of the MBGs properties with those of galaxies found in other surveys 
(Coziol\markcite{c9} {\em et al.} 1994, hereafter Paper~II) suggested that
our survey is biased towards the low-excitation Starburst Nucleus Galaxies (SBNGs). 
In general, the SBNGs are more massive and more chemically evolved than the high 
excitation HII galaxies (Salzer {\it et al.} 1989; Terlevich {\it et al.} 1991). 
 
The mechanism responsible for the starburst phenomenon is still unknown 
(see Gallagher\markcite{g1} [1993] for a brief discussion of possible alternatives). 
For the massive SBNGs, it is usually believed that they are evolved galaxies which were 
rejuvenated by a recent infall of matter (Huchra 1977). 
It is also believed that this new injection of matter is
related to some kind of interaction with another galaxy.  
To verify this hypothesis, we have taken CCD images of a sub--sample of the MBGs 
(Coziol {\em et al.} 1995; Barth\markcite{b4} {\em et al.} 1995). 
Although many of these galaxies are clear examples 
of interacting or merging galaxies, almost half are relatively isolated. 
A more complete analysis of our images further revealed that all the isolated 
galaxies present peculiar morphological characteristics that could 
be interpreted as indirect signs of interactions. 
These galaxies could be remnants
or advanced stages of past interacting galaxies (Barth\markcite{b4} {\em et al.} 1995).   
But this would also imply that 
the present burst of star formation in these galaxies is somehow decoupled from 
the presumed triggering event (Coziol\markcite{c7} {\em et al.} 1995).   
Confirmation of this behavior for a larger number
of SBNGs would suggest that in some of the massive starbursts the star formation process
may depend on internal as well as external causes.
   
Our third addition increases by $\sim 40\%$  the 
number of MBGs. This allows us to better define the global properties of
our candidates and verify some of the results presented in
our previous papers. The plan of this article is the following.
The third list is described in section~2. 
The results of our last spectrophotometry follow--up are presented in section~3. 
New photometric observations of bright candidates are described in section~4. 
We rediscuss the completeness limit of our survey in section~4.
In section 5, we conclude with a brief discussion on the possible
causes at the origin of the bursts in our candidates.  

\section{THE THIRD LIST OF MBG CANDIDATES}

The third list of 135 MBG candidates ensues from the inspection of 
a set of 59 plates, corresponding to an area of 1376 deg$^2$.
This third installment brings the level of completion of our survey to 63\%, 
and reaches now 4400 deg$^2$. 
Information on the new candidates identified is compiled in Table 1.
It follows the same format used in the previously published tables.  
The name of the object is based on its 1950 equatorial coordinates.  
The calibrated photographic B magnitudes and U$-$B colors are obtained from the APM measurements. 
As determined in  Paper~I, the mean uncertainty of the B$_{APM}$ is $\pm$ 1 magnitude.
It is important to remember that the machine determined U$-$B color 
is not a reliable indication of the true color (see Paper~I). 
It is included in our list only for the sake of completeness.
Absolute magnitudes are estimated using the B$_{APM}$ and the radial 
velocity given by the NASA/IPAC Extragalactic Database (NED), assuming
H$_\circ$ = 75 km s$^{-1}$ Mpc$^{-1}$.  
Uncertainties for the coordinates are of the order of arcseconds.
Other information found in Table 1 are cross-identification with objects found in NED.   
An ``n'' in the last column refers to a note in the Appendix.  
In this third list, several galaxies were not found in the APM file of
the particular Schmidt plate. Because of some peculiarities in their images, 
the APM machine ignored those objects. We include them in our list, 
because they represent good UV-bright candidates.
In Table 1, those few  cases are identified by a ``*''.
For these galaxies, we quote the B magnitude and coordinates as found in NED. 

The total list of MBGs amounts to 469 galaxies brighter than B$_{APM}$ = 15.5. 
It corresponds to a density of 0.11 galaxy per deg$^2$. This is almost 
the density of galaxies found by the Markarian survey (Mazzarella \& Balzano 1986). 
By comparison, the densities of galaxies found in
the Kiso (Comte {\it et al.} 1994) and the Northern 
Case surveys (Salzer {\it et al.} 1995) represent  
15 times and 13 times the density per deg$^2$ of the MBG survey.
These differences are explained by the fact that
these two surveys go at least one magnitude deeper than the MBG survey (B $\sim 16.5$). 

A fraction of the fields covered by the Kiso survey 
are located south of the celestial equator. 
Because the Kiso magnitudes are roughly 
similar to those of the APM, we can arbitrarily assign a magnitude limit to this
survey and compare some of their output to our results. 
Using the lists of Takase \& Miyauchi-Isobe (1993), 
we determined that 96 Kiso UV-bright galaxies (or KUGs), 
with an apparent magnitude brighter or equal to
B = 15.5, are located in the area overlapping with our 187 fields. 
Only $\sim 20$\% of these galaxies are included in our 3 lists.
We note that in the Kiso survey the degree 
of UV-excess of the KUGs is coded H, M and L, corresponding to high, medium and low. 
The MBGs include 12 out of the 66 (18\%) KUGs marked L , 5 out of the 24 (21\%) marked M,  
and 5 out of the 6 (83\%) marked H. Contrary to the Kiso survey, we do not try to establish 
any gradation in the level of UV-excess of our candidates. 
Therefore, the MBG Survey detects essentially all of the strongest UV-excess
objects, but relatively few of the intermediate and low
UV-excess galaxies. This comparison suggests that the MBG selection
criterion is more restrictive than that used by the Kiso
survey.

\section{SPECTROPHOTOMETRY}

\subsection{\it Data acquisition}

Spectrophotometric observations of a sample of 22 galaxies were 
obtained with the 2.1 m telescope located at the Observatorio Astron\'omico Nacional, 
San Pedro M\'artir, B. C., M\'exico, during 4 nights in October 1993. 
The configuration of the instruments was exactly as described in Paper~II. 
A Boller \& Chivens spectrograph, equipped with a blue coated, $1024 \times 1024$
Thompson CCD detector was employed. We used two gratings of 600 lines/mm 
(blazes 13 and 8.63 degrees) to obtain red (5800--7200 \AA) and blue (4400--5900 \AA)
spectra of the candidates. All the observations were made with the slit centered on
the nucleus of the galaxy, oriented E-W, and a slit width of 150 and
250 $\mu$m, corresponding to 2.0 and 3.3 arcseconds in the sky. For
each galaxy, we obtained two spectra in the red (10 min. exposure time)
and two in the blue (20 min. exposure time). The mean resolution is 4
\AA\ in the red and 8 \AA\  in the blue. A He-Ar lamp was used for wavelength calibration.
Flux calibration was provided by additional observations of standard stars from the
lists of Stone\markcite{s8} (1977), Stone\markcite{s9} \& Baldwin (1983) 
and Baldwin \& Stone\markcite{b2} (1984). 
Due to poor weather conditions, the fluxes measured during
the first night of observation are poorly determined. Uncertainties
of the fluxes for this night are higher than 20\%.  For data
reduction, we have followed the same procedures as described in
Paper~I and Paper~II. 

\subsection{\it Data analysis} 

One of our goals during our most recent observing run at San Pedro M\'artir was to complete
the observations made during the previous missions. We therefore 
reobserved in the blue almost all of the galaxies for which we already had
a red spectrum. Only 6 new galaxies were observed in the red. The
results are presented in Tables 2 and 3. In Table 2, the
intensities of the lines are given relative to H$\alpha$. In Table 3
they are given relative to H$\beta$. As in the two previous papers, 
the lines were measured by adjusting gaussian fits to the profiles,
setting the continuum by eyes. No correction for galactic
interstellar reddening or intrinsic reddening has been applied. The
uncertainties on the fluxes are of the order of 10\%, and represent the internal
consistency of our method only. It was determined by comparing the values
measured on two different (independent) spectra of the object. 
In Table 2, the radial velocities were derived from the average of the detected lines, 
but were not corrected for the Earth motions. Uncertainties for the radial velocities
are of the order of 60 km s$^{-1}$. 
Also listed in Tables 2 and 3 are the full width at half maximum (FWHM),
corrected for the instrument profile, and the equivalent width (EW)
of H$\alpha$ and H$\beta$. 

In Fig. 1, we show the spectra of MBG22342-2228 which was classified 
in Paper~II as a possible AGN. 
The additional blue spectrum allows us to classify this galaxy as a Seyfert~1.   
Fig. 2 shows the available red spectrum of MBG03536-1351. 
The [NII]$\lambda$6584/H$\alpha$ ratio is high suggesting that it could be another AGN.
Based on the absence of a wide component for the H$\alpha$ line, we tentatively 
classified this galaxy as a Seyfert 2. Observation in the blue part of the spectrum will
be necessary to confirm this classification. 
Our current yield of AGNs implies that barely 5\% of our galaxies will be of this nature.  
 
As in Paper~II, we have estimated the metallicity of our candidates using the
diagnostic diagram [OIII]$\lambda$5007/H$\beta$ vs. [NII]$\lambda$6584/H$\alpha$. 
Our method is based on the calibration of this diagram using
a sample of galaxies from the literature for which reliable chemical 
abundances exist (see Paper~II).
The uncertainty associated with our method is approximately 0.3 dex.
The estimated metallicities for 19 MBGs are presented in Table 4. 
In this table, we give also the electron densities of some of the galaxies.
These densities were estimated using the sulfur doublet [SII]$\lambda\lambda$6716,6717, assuming
for the ionized gas an electron temperature of 10 000 K. Both the
metallicities and electron densities are typical of giant extragalactic HII regions
located in the central regions of spiral galaxies.

\section{NEW PHOTOMETRY AND COMPLETENESS LIMIT} 

In Paper~II of this series we used the V/V$_m$ test (Schmidt\markcite{s5} 1968; 
Sargent\markcite{s10} 1972; see Paper~II)
to establish the completeness of our survey. It was then realized that 
for bright galaxies, the magnitude given by the APM is often far from the one
obtained by luminosity profile fitting (see Fig. 1 of Paper~I).
To palliate this problem we have therefore initiated a program
of CCD photometry of the brightest MBGs without accurate 
magnitude in NED. Galaxies with B$_{APM} < 14.0$ which can be observed
from the north latitude were selected.
The B,V CCD photometry was acquired at the 1.6 m telescope
of the Mont M\'egantic, using a THX Thompson $1024\times 1024$
CCD camera. The data were obtained during five nights in October 1993. 
On each night a number of
Landolt's\markcite{l2} (1992) equatorial standards were concurrently observed to 
determined the extinction and transformation coefficients. Details of the
observations and data reduction can be found in Barn\'eoud\markcite{b6} (1994). Total
galaxy magnitudes were determined using IRAF aperture photometry package.
Table 5 lists the total B and V magnitudes integrated up to the sky
level which is reached asymptotically. R$_V$ is the adopted radius,
in the V band, where the sky is reached. The uncertainties on B$_T$ and
V$_T$ are estimated to be 0.08 and 0.07 respectively. The major
contributor to this uncertainty comes from the adoption of the value
of the sky brightness (Barn\'eoud 1994). The last column gives the B--V colors
of the galaxies. Nearly all B--V colors are between 0.40 and 0.75. This agrees 
with the color distribution of non Seyfert Markarian galaxies published
by Huchra (1977). 

With better estimates of the apparent magnitudes at the bright end,
and an increase by 40\% in the number of MBGs, we take a second look at the
completeness of our survey using the $<V/V_m>$ method.
For this test, we used the best estimates of B magnitudes of
galaxies: these include 41 galaxies with CCD photometry, 152 galaxies with
photoelectric magnitudes and 235 galaxies with APM measured magnitudes.
The new results of the $V/V_m$ test are listed in Table 6. 
The first column gives the absolute magnitude of each interval considered, the second 
column gives the cumulative number of galaxies 
up to the absolute magnitude of the interval and the last column gives 
the cumulative number of galaxies to be added to correct for the 
incompleteness up to this magnitude.
The results are shown in Fig. 3.  
The error bars correspond to $\sigma = (12N)^{-1/2}$, where N is the number
of galaxies (Green\markcite{g2} 1980).
This new test suggests that our survey is complete to B = 14.7 and 
more than 90\% complete at B = 15.0.  
 
\section{THE NATURE OF THE MBGS AND THE ORIGIN OF THE BURSTS}

\subsection{\it Spectroscopic characteristics, redshifts and infrared luminosities}

With the addition of new spectra, we double the number of MBGs
that can be classified using diagnostic diagrams. 
In Fig.~4, we compare the spectral characteristics of our candidates  
with those of emission--line galaxies from three other samples: the
catalogue of HII galaxies (CHIIG; Terlevich\markcite{t3} {\em et al.} 1991), 
a sample of luminous IRAS galaxies 
(Allen\markcite{r1} {\em et al.} 1991) and a sample of compact 
Kiso galaxies (Comte\markcite{c4} {\em et al.} 1994). 
Fig.~4 illustrates clearly the biases between the different samples.  
Establishing the boundary between low and high excitation galaxies
at Log([OIII]$\lambda$5007/H$\beta) = 0.4$ (Coziol\markcite{c6} 1996),  
our new data confirm that almost all of the MBGs, 
and all the IRAS galaxies, are low--excitation SBNGs. 
The sample of compact Kiso survey seem to contains a slightly higher
fraction of HII galaxies.
Similarly, Salzer {\em et al.} (1995) shown that 
the fraction of low--excitation starbursts detected in the Case survey is 
also higher than usually found by other prism-objective surveys.
These comparison suggests that samples of starburst galaxies suffer various biases  
which mainly reflect the combination effect of selection criteria and of the magnitude limits
reached by the different surveys. This in particular could
explain the bias of the MBG survey towards the SBNGs.
In general, HII galaxies have strong emission--lines and high UV-excess, but a relatively faint
continuum in B as compared to the SBNGs (Coziol 1996). This is because the SBNGs
are located in more massive galaxies with a higher fraction of
intermediate aged stellar populations, which dominate in the B band.  
Because we select our candidates by eye our magnitude limit
is relatively bright and we are selecting preferentially 
galaxies which are more extended and luminous in B.  
For the same reason, by using a mixed selection criterion, the Case survey is 
able to detect a higher fraction of low--excitation SBNGs than it would   
if it was selecting its candidates only based on the presence of emission--lines.   
Consequently, the deeper magnitude limit reached by the Kiso survey, coupled to its extended definition
of UV-excess galaxies allows it to detect a higher fraction of HII galaxies. 

In Fig.~4, the continuous curve represents the mean position of standard HII regions.
We are using this curve to calibrate our diagnostic diagram in terms of metallicities 
(see Paper~II). The working principle behind our method is
explained by the different models of HII regions, which show that the main
parameter reproducing this sequence is a variation of the metallicities
(McCall {\it et al.} 1985, hereafter MRS; Dopita \& Evans 1986).
MRS (1985) have also shown that normal HII regions in the disk of spirals
trace a tight distribution around this curve (see their Fig.~3).  
This could be explained if in general HII regions are
ionization--bounded (MRS 1985). In comparison, we can see in Fig.~4
that both the MBGs and the IRAS starbursts 
are systematically located to the right of our calibration curve.
On average, the SBNGs show an excess emission of [NII]$\lambda$6584
as compared to normal HII regions. Surprisingly, this phenomenon is not obvious  
for the compact Kiso galaxies. 
At the bottom of Fig.~4, we have put the error bars corresponding 
to 20\% uncertainties in the line ratios. 
Although such uncertainty is compatible with the
dispersion of the [NII]$\lambda$6584/H$\alpha$ ratios, 
it cannot explain the systematic trend towards higher values.
To quantify this phenomenon, we increased by 0.2 dex 
the ratio [NII]$\lambda$6584/H$\alpha$ as predicted by the MRS model. 
In Fig.~4, the good fit between the shifted MRS model and the
data indicates a mean excess of 0.2 dex in
the [NII]$\lambda$6584/H$\alpha$ ratio as compared to normal HII regions.  

Excess emission of [NII]$\lambda$6584 was already observed before in other samples of
emission-line galaxies. It seems like a very common phenomenon in the nucleus of 
``normal'' galaxies (Stauffer\markcite{s7} 1982). 
Our observations show now that this is also a characteristic of SBNGs.
At the moment, nothing allows us to determine the
cause of this phenomenon in the SBNGs. 
In the literature, various hypotheses have already been suggested.
For example, this excess of emission could indicate an overabundance of Nitrogen
in the nuclei of galaxies (Stauffer 1982). 
The same solution was proposed to explain the typical 
high ratios of [NII]$\lambda$6584 in AGNs (Storchi--Bergmann \& Pastoriza 1989; 
Storchi--Bergmann 1991). Considering the possible relation
between starburst and AGNs, it would be interesting to know if  
this overabundance of Nitrogen could have the same origin in both types of galaxies? 
But this excess of emission could also imply a supplementary source of
ionization. All these galaxies may contain an important 
quantity of hot diffuse gas, which was either  
shock-ionized by supernovae and stellar winds (Lehnert\markcite{l1} \& Heckman 1994), or 
even excited by an unresolved weak AGN hidden in their nuclei 
(Kennicutt\markcite{k3} {\em et al.} 1989). Note that these two possibilities
are very much similar to the alternatives proposed to explain the nature of LINERs. 
One could finally suppose that the excess emission of Nitrogen 
is solely due to selective depletion of cooling elements 
during dust formation (Shields\markcite{s4} \& Kennicutt 1995). 
In principle, all these hypothesis could apply to the SBNGs. 
Depending on the solution however, the nature of the SBNGs
would be very much different.  
The exact cause of the excess emission 
of Nitrogen in SBNGs seems therefore essential to establish
if we want to understand the nature of these objects.
The solution to this problem
would probably also help us understand what kind of relation 
could exist between starbursts and AGNs.

The similarities between the spectral characteristics of the MBGs and 
of the luminous IRAS galaxies suggest that these
galaxies have a common nature. 
In Fig.~5, we compare the redshifts and far--infrared luminosities 
of the MBGs with those of luminous IRAS galaxies.  
The sample of IRAS galaxies is composed of the galaxies from the sample of 
Allen\markcite{r1} {\em et al.} (1991) and the sample of luminous infrared galaxies 
from Veilleux\markcite{v2} {\em et al.} (1995). 
In general, the MBGs possess lower infrared luminosities 
(L$_{IR} < 10^{11}$ L$_{\odot}$)
and are nearer (z $< 0.05$) than the luminous IRAS galaxies. 
The lower far--infrared luminosities of the MBGs 
suggests that the MBGs could be starburst galaxies at a different 
stage of evolution than the luminous IRAS galaxies.

\subsection{\it Morphologies of the SBNGs}

In Fig.~6, we present the distribution of the
morphologies of the MBGs. Information on the morphologies come 
from NED or are based on our own CCD imaging (Barn\'eoud 1994; Barth {\em et al.} 1995). 
Only 39\% (182) of the MBGs are morphologically classified.
Fig. 7shows that a high fraction of the MBGs are early-type spirals (Sb and earlier). 
This characteristic is not unique to our sample.
In Fig.~6, we show also the distribution of the morphologies of 
the Markarian galaxies (Mazzarella\markcite{m5} \& Balzano 1986).
In this list, the fraction of galaxies which 
have their morphology classified is similar to ours, although the
number of galaxies is almost double (38\% or 349 galaxies).   
The distribution of morphologies of the MBGs and the Markarian galaxies are nearly identical.
In Fig.~6, we included also the distribution of the morphologies of the SBNGs as defined
by Balzano (1983). As compared to the two other samples, this sample 
contains a lower fraction of early--type spiral galaxies (earlier than Sa). 
In paper~II, we verified that no difference exists in the intensity of star formation  
of the early-type MBGs as compared to those 
of the the starbursts in Balzano's sample. 
The observed bias towards the late-type spirals in this last sample 
can only be explained by the particular selection criteria used 
by this author to define the SBNGs.
From our comparison, it is suggested that 
we should extend the definition of SBNGs in order to include a
larger fraction of early--type spirals.   

\subsection{\it The roles of bars and interactions in the SBNGs}

The presence of a nonaxisymetric feature, like a bar, is frequently
suggested to explain a nuclear starburst.   
Despite the many efforts devoted to this subject however, the current observational 
evidences are still ambiguous and controversial.  
For example, using a complete sample of infrared luminous galaxies, 
Devereux\markcite{d2} (1994) found that it is
in the early-type barred spirals that the nuclear star formation rate is enhanced. 
This result was contradicted later by Giuricin {\it et al.} (1994) 
who suggested that the effect seen by Devereux simply reflects
the fact that in early-type spirals the
emission is usually more compact than in late-type ones. These authors have also found that it is 
the late-type barred galaxies that show an enhancement of star formation in their nuclei.
In a recent paper, Martin (1995) found that  
71\% of the galaxies with a nuclear starburst have a strong bar structure, in comparison
with 59\% for the quiescent galaxies. A quick examination of the Martin's sample 
clearly show the preponderance of late--type spirals. We may therefore 
conclude that it is the late--type starburst that are predominantly barred.

In Fig.~6, we distinguished between galaxies with 
and without a bar. Note that for these statistics we considered
SAB as barred. About 35\% of the Markarian galaxies and 48\% of the MBGs have a bar. 
This proportion is slightly higher in Balzano's sample, with 58\% of barred galaxies.
In Fig.~6, it is clearly observed that the frequency
of bars detected depends on the morphology of the host galaxy, and
increases towards late-type spirals. The fact that  
Balzano's sample is biased towards the late--type
spirals explained therefore the higher fraction of barred galaxies in this sample.
With the limited information now available for our candidates, we cannot say if the
variation of the frequency of bar detection as a function of the morphology 
corresponds to a real effect, or
if it means that a bar is more difficult to detect when the galaxy has a strong bulge.   
Deeper CCD observations or H$\alpha$ imaging
of the MBGs will be required to settle this question.  

To push our analysis further, we determined the frequency of isolated MBGs with a bar.  
For this test, we distinguished between early-type starbursts (Sb and earlier) from the 
late--type starbursts (all spiral types later than Sb). 
Only 37\% of the isolated
early-type starbursts have a bar, as compared to 61\% for the isolated late--type ones. 
Note that among the early-type starbursts without a bar, 37\% are of type E and SO, as compared
to 24\% for the Hubble types Sa, Sab and Sb. This suggests that 
the observational bias effect is marginal. 
Isolated late-type spirals show a clear tendency to be barred. This
result is consistent with those of Giuricin {\it et al.} (1994) and Martin (1995).

Another popular idea is that massive starbursts occur preferentially
within galaxies undergoing interactions.
For all the MBGs, we determined the frequency of interacting galaxies by 
visual inspection of the Palomar and the ESO/SERC atlases. 
Only 24\% (111) of the galaxies in our sample show a clear evidence 
of interaction. Obviously, this kind of research is limited by the 
resolution of the plates used. More complete analysis of CCD imaging  
will be require to detect weak signals of interaction (see Barth {\it et al.} 1995).    
Our estimated frequency of interacting MBGs is therefore only a lower limit. 
This limit is already sufficiently low however to suggest that 
in a high number of massive galaxies the burst of star formation 
is probably not solely caused by dynamical processes.  
In a recent paper, based on a study of HII galaxies,
Telles\markcite{t2} \& Terlevich (1995) arrived at exactly this conclusion. 
According to these authors, only 21\% of the HII galaxies show clear signs of interaction. 
For small mass galaxies, it was suggested that the bursts in the HII galaxies could 
come from interactions with invisible HI companions (Taylor\markcite{t5} {\em et al.} 1993). 
At the moment however, this hypothesis is far from being 
verified (see Taylor\markcite{t4} {\em et al.} 1996). 
Furthermore, for the more massive SBNGs this alternative is even less acceptable.  
Indeed, it would be difficult to explain the existence of massive HI 
companions near massive galaxies. 
Therefore, the tendency towards relative isolation, as suggested 
by our statistics for the MBGs, should constitute a
strong argument in support of the idea that the starburst phenomenon 
in massive galaxies may also depends on internal mechanisms. 

\section{SUMMARY AND CONCLUSION}

This third installment of the MBG survey brings it to  63\% completion.
The total list of MBGs amounts now to 469 galaxies brighter than B$_{APM}$ = 15.5.
New result of the $V/V_m$ test shows that our survey is complete to B = 14.7.
Our new spectroscopic informations confirm also the bias of our survey towards the
low-excitation and metal rich SBNGs. 
The distribution of morphologies of the MBGs is similar to the one for the
Markarian survey and indicates that SBNGs are very common among 
the early-type spirals (Sb and earlier).

The spectral characteristics of our candidates are similar 
to those of the luminous IRAS galaxies.
But the MBGs are generally nearer (z $< 0.05$) 
and possess lower infrared luminosities (L$_{IR} < 10^{11}$ L$_{\odot}$) than 
these galaxies. View the lower infrared luminosities of the MBGs, 
these galaxies could be starburst galaxies at a different stage 
of evolution than the luminous IRAS galaxies.

As a common feature with the luminous IRAS galaxies, the MBGs  
show on average an excess emission of Nitrogen of 0.2 dex as compared to 
normal disk HII regions. Different mechanisms could produce this effect in the SBNGs and
at the moment nothing could allow us to distinguish between the different alternatives. 
The determination of the excat origin of the excess emission of Nitrogen
in SBNGs is essential if we want to understand the nature of these galaxies. The solution
of this problem could probably also help understand what kind of relation could
exist between starburst galaxies and AGNs.

Based on the available informations, we are unable to identify
one mechanism responsible for the burst of star formation in all the MBGs. 
This goes contrary to the opinion that starburst in massive galaxies 
are all produced by interaction. 
Obvious interacting galaxies constitute only 24\% of our candidates.
Although this number is only a lower limit, it is sufficiently 
low to suggest that in some massive galaxies the burst of star formation may depends on  
internal mechanisms independent of any interaction.

Such internal mechanisms could perhaps be related to a bar.
If we consider only the isolated MBGs, we find that
the late-type spirals show a clear tendency to be barred.
But in our sample, the frequency of bar detection also depends on the 
morphology of the host galaxy and increases towards the late-type spirals. 
We cannot tell if this is a real effect or if it is due to the difficulty of
detecting bars in early-type galaxies.  

Perhaps the origin of the burst is different in 
starburst galaxies of different morphological types.
Bars could be responsible
for the enhanced star formation in the nucleus of the late--type spirals, and
interactions and mergers could be responsible for the bursts
in the early-type galaxies. In these last cases, the bursts
should also depend on some self--regulated mechanism, whish is necessary
to explain the relative isolation of these galaxies. 

\acknowledgments

RC acknowledge the financial support of the Brazilian 
FAPESP ({\em Funda\c{c}\~ao de Amparo \`a
Pesquisea do Estado de S\~ao Paulo}), under contract 94/3005--0.
The financial support of the Natural Sciences
and Engineering Research Council of Canada and the Fonds FCAR du Qu\'ebec
are also gratefully acknowledged.    
The authors acknowledges the anonymous referee 
for his comments which have considerably contributed to improve the paper.
This research has made use of the NASA/IPAC Extragalactic Database
(NED) which is operated by the Jet Propulsion Laboratory, California
Institute of Technology, under contract with the National Aeronautics 
and Space Administration. 

\newpage
\appendix 
\section{Notes on individual objects}
\begin{description}
\item[00485-0719] Known interacting galaxy, ARP 140.
\item[01056-0449]In a group which includes IC 76 and MRK 973.
\item[01137-5027]Known Sy2 galaxy.
\item[01578-6806]Known emission-line galaxy (Bettoni\markcite{b5} \& Buson 1987).
\item[02043-5525]Known Sy2 galaxy.
\item[03019-2615]This galaxy is not as bright as the B$_{APM}$ would
suggest. The total magnitude listed in NED is B$_T = 11.78$. The M$_B$
quoted here, calculated with B = 8.3, is thus overevaluated. 
\item[03023-2739]This galaxy is also known as HARO 19.
\item[03027-2742]Galaxy pair, the given B$_{APM}$ is just an eye estimate.
No magnitude is given in NED. 
\item[03540-4229]Arp-Madore interacting galaxy (Sekiguchi\markcite{s2} \& Wolstencroft 1993).
\item[04019-4332]Known Wolf-Rayet galaxy (Conti\markcite{c5} 1991).
\item[04163-5017]Known Seyfert.
\item[04350-4017]Galaxy pair.
\item[21068-3742]The listed magnitude in NED is B$_T = 13.78$, the
magnitude obtained by the APM does not correspond to the whole galaxy. We
keep this galaxy because it is obviously brighter than B = 17.
\item[21397-5255]Galaxy pair.
\item[22543-3643]The quoted magnitude in NED is B$_T = 10.97$. Our 
measured magnitude is too bright. The listed M$_B$ is also too bright.
\item[22551-3747]Galaxy pair.
\item[22565-4809]Galaxy pair.
\item[22566-3758]The quoted magnitude in NED is B$_T = 11.84$, the measured
magnitude, by the APM does not refer to the whole galaxy. Therefore, our
calculated M$_B$ is too faint. 
\item[2335-3815]The quoted magnitude in NED is B$_T = 11.51$, the measured
magnitude by the APM does not refer to the whole galaxy. Therefore, our
calculated M$_B$ is too faint.
\item[23391-3654] Galaxy pair.
\end{description}

\newpage
\figcaption[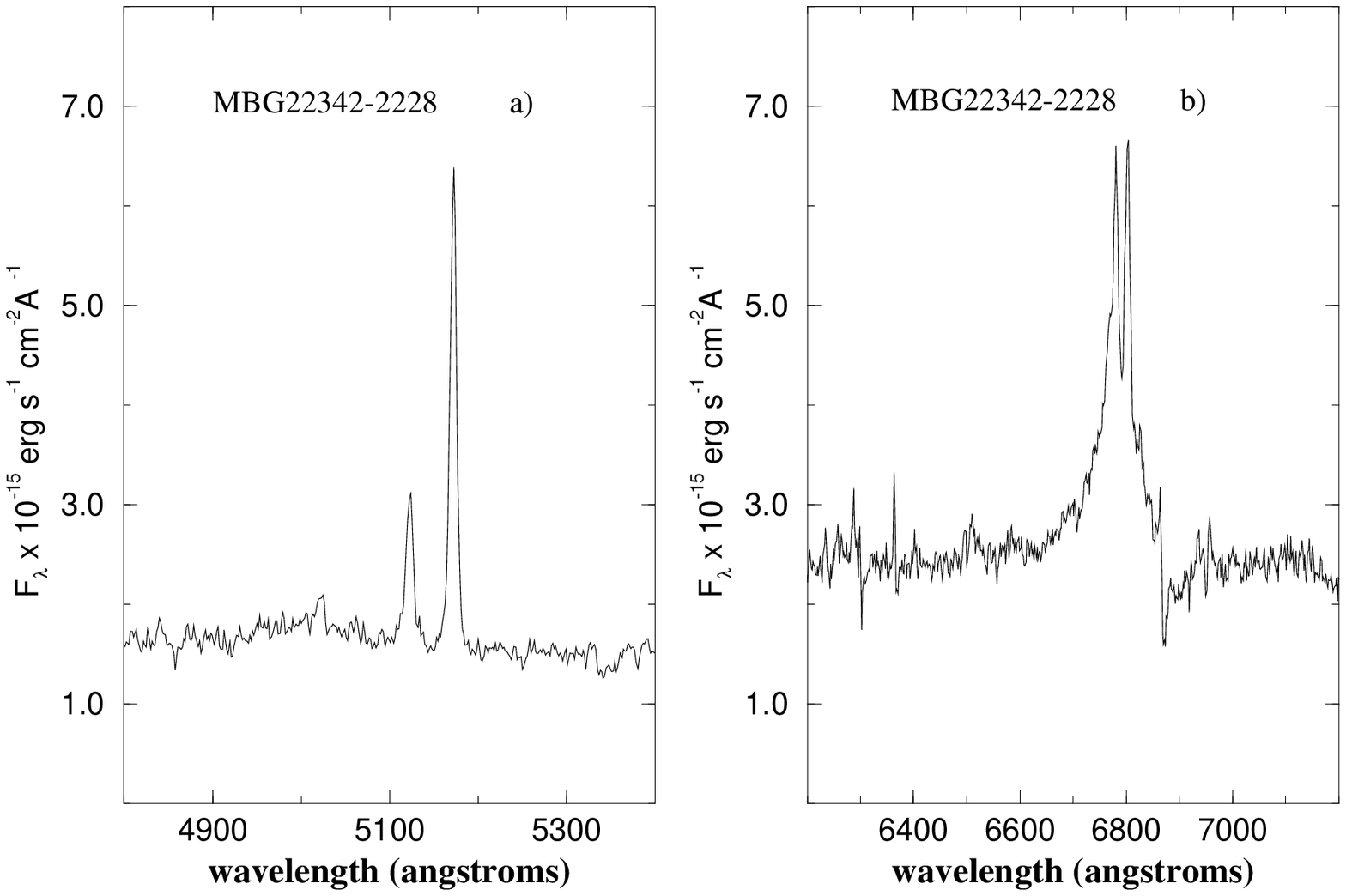]{Spectra of the galaxy MBG22342-2228. 
a) Blue part of the spectrum centered on H$\beta$.
b) Red part of the spectrum centered on H$\alpha$. The wide component of the Balmer 
lines implies that it is a Seyfert 1.}

\figcaption[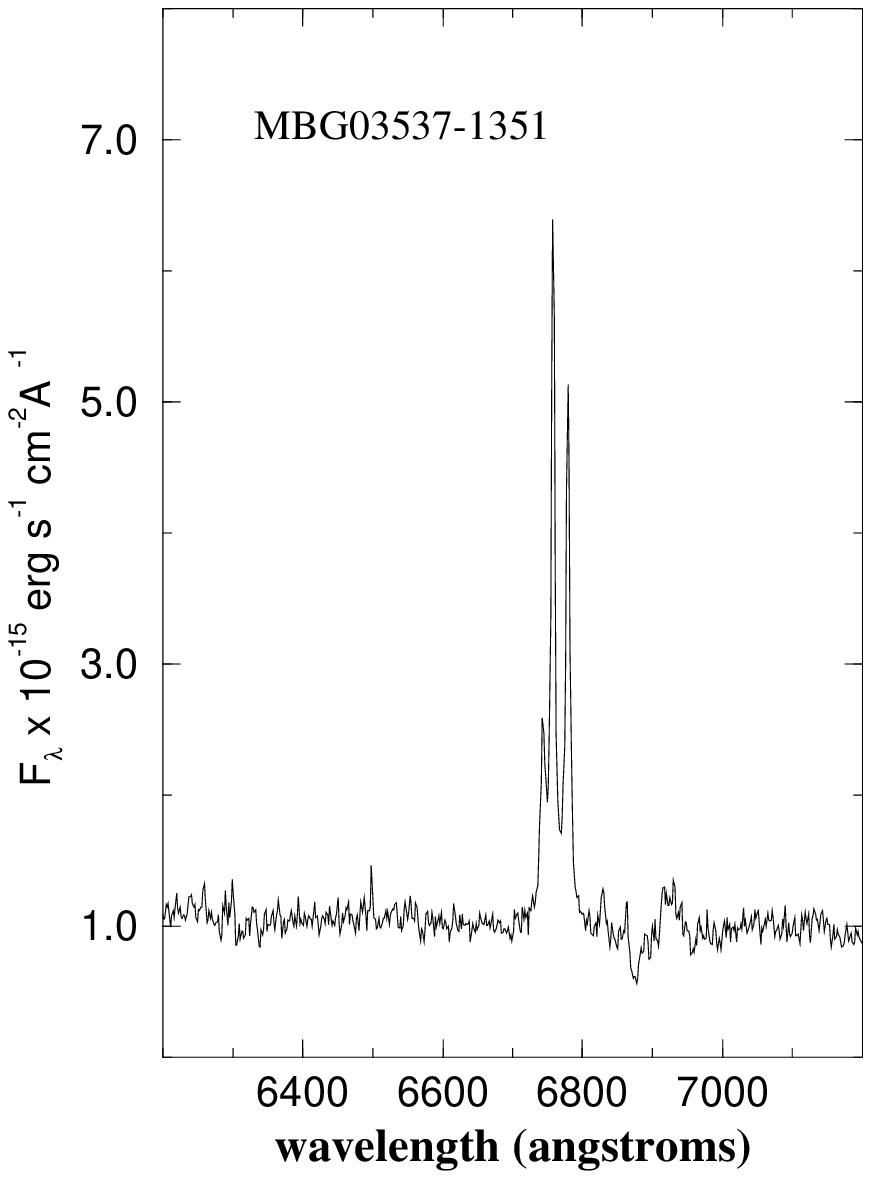]{Red spectrum centered on H$\alpha$ of a possible new Seyfert 
galaxy among our candidates. Based on the absence of obvious large components 
of the Balmer line, we tentatively classify this galaxy as a Seyfert 2.}

\figcaption[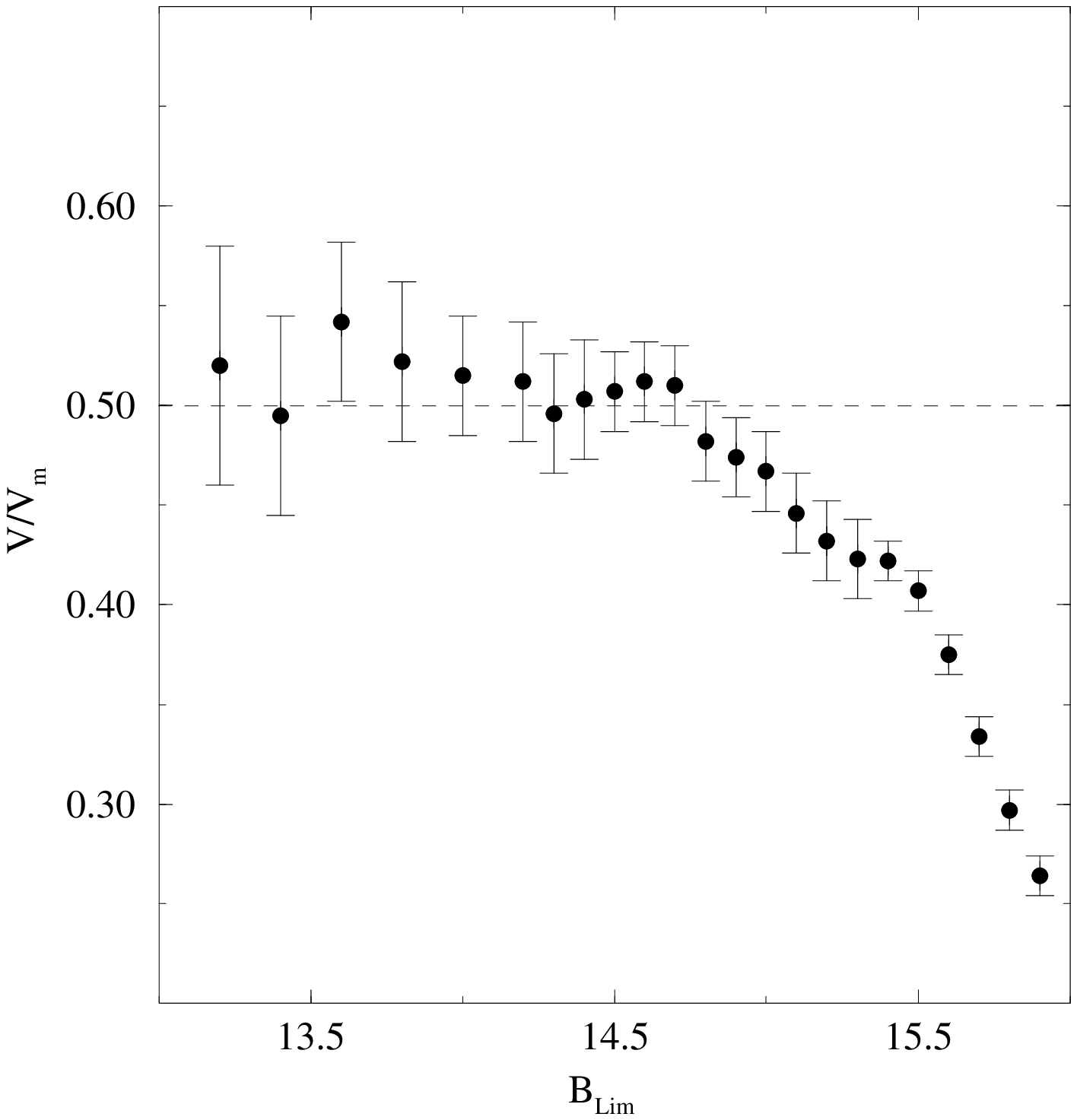]{New V/V$_m$ test for the MBGs. The MBG survey is 
complete to B$ = 14.7$}

\figcaption[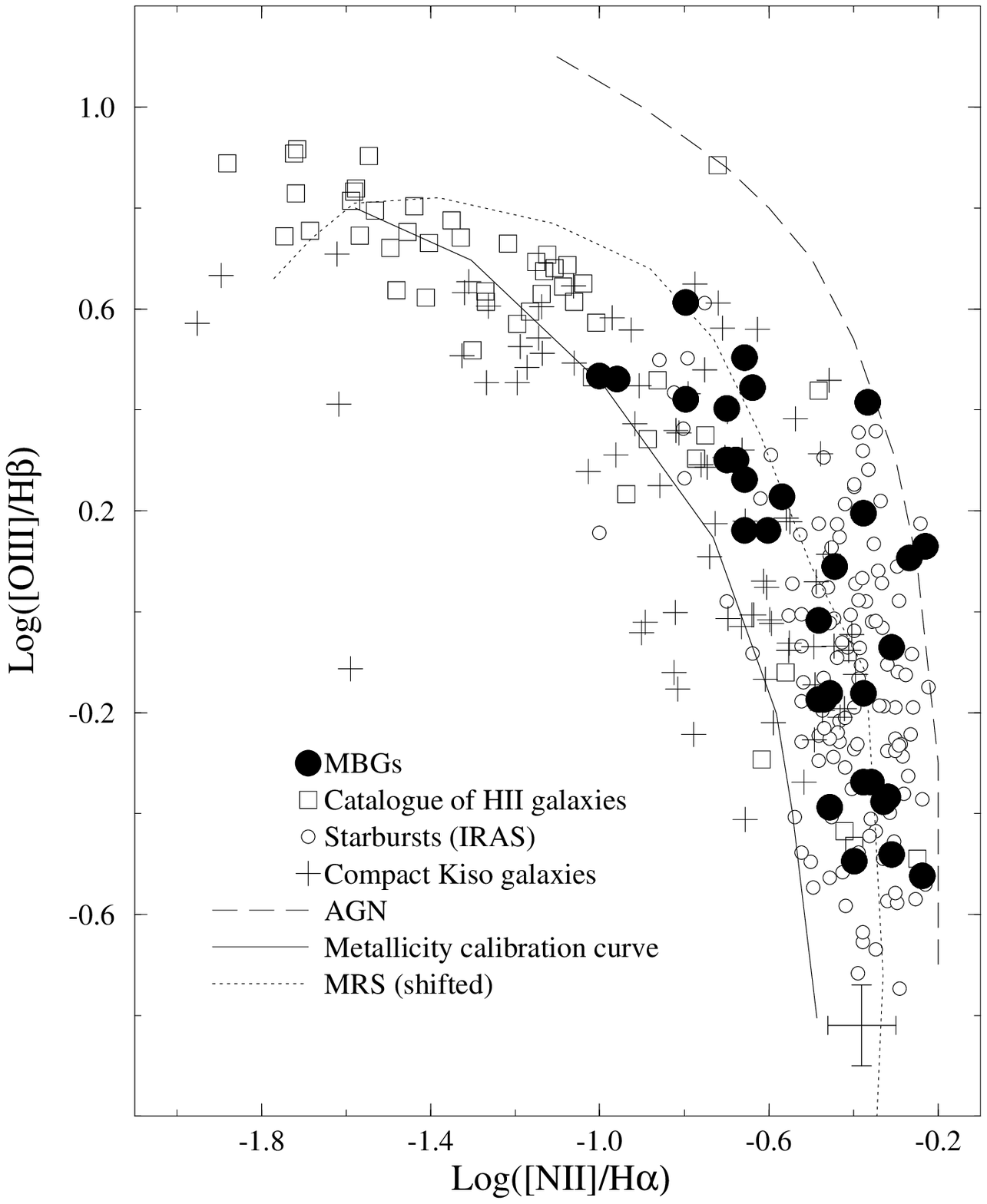]{Diagnostic diagram of [OIII]$\lambda$5007/H$\beta$ vs. 
[NII]$\lambda$6584/H$\alpha$. The long--dashed line is the empirical 
separation between starbursts and AGNs (Veilleux \& Osterbrock 1987). The solid line is the 
metallicity calibration curve as determined by Coziol {\em et al.} 1994. 
The model of MRS (dotted line) was shifted to fit the data for 
the SBNGs: there is an excess of 0.2 dex in the ratio [NII]$\lambda$6584/H$\alpha$
of the SBNGs. The error bars correspond to uncertainties of 20\% on the line ratios.}

\figcaption[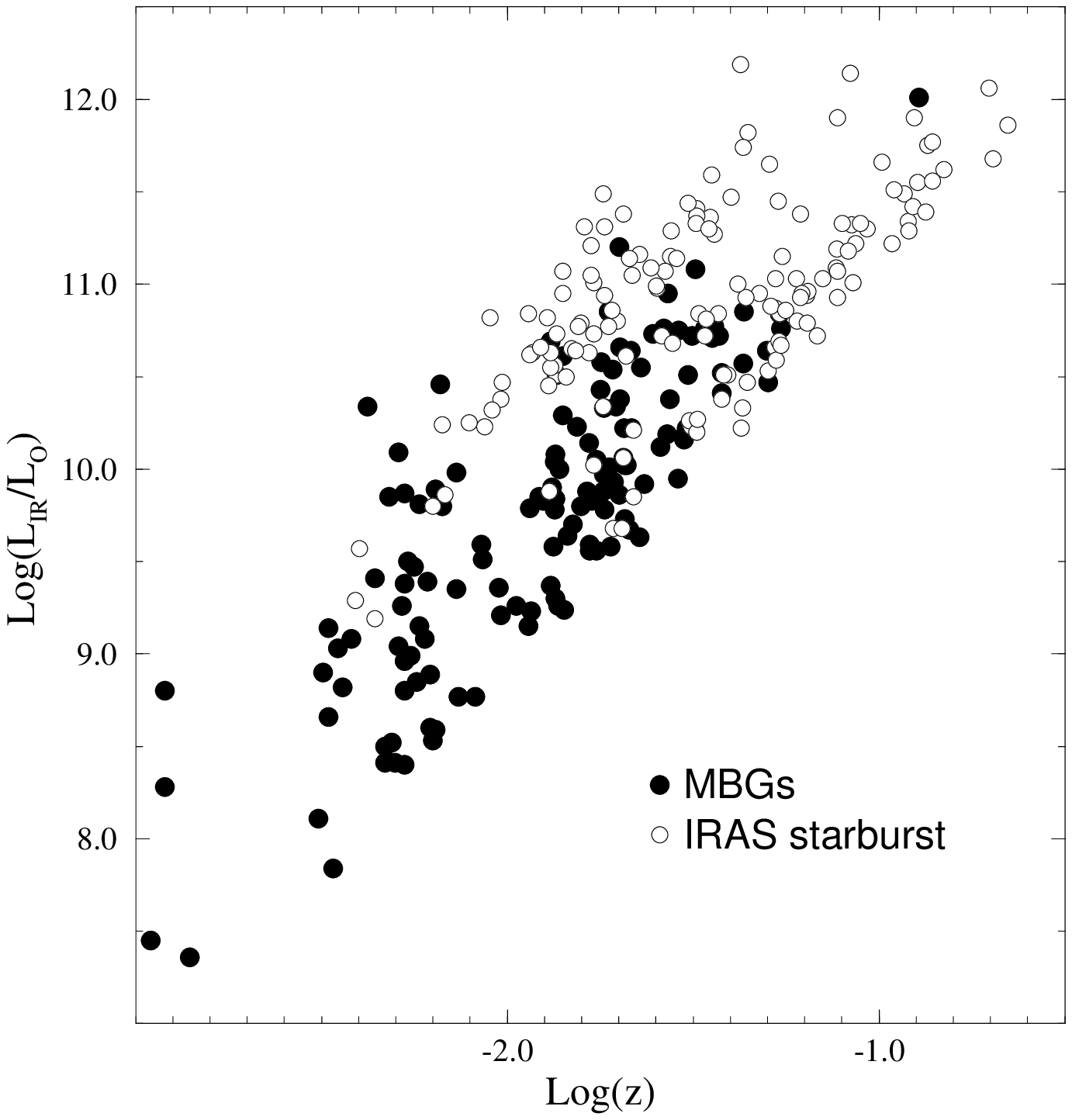]{Diagram of the far--infrared luminosities as a function
of the redshift. The MBGs are compared to the IRAS starburst from Allen {\em et al.} (1991), 
and the luminous infrared galaxies from Veilleux {\em et al.} (1995). The MBGs are 
less luminous in infrared and closer than the galaxies of these two samples.}  

\figcaption[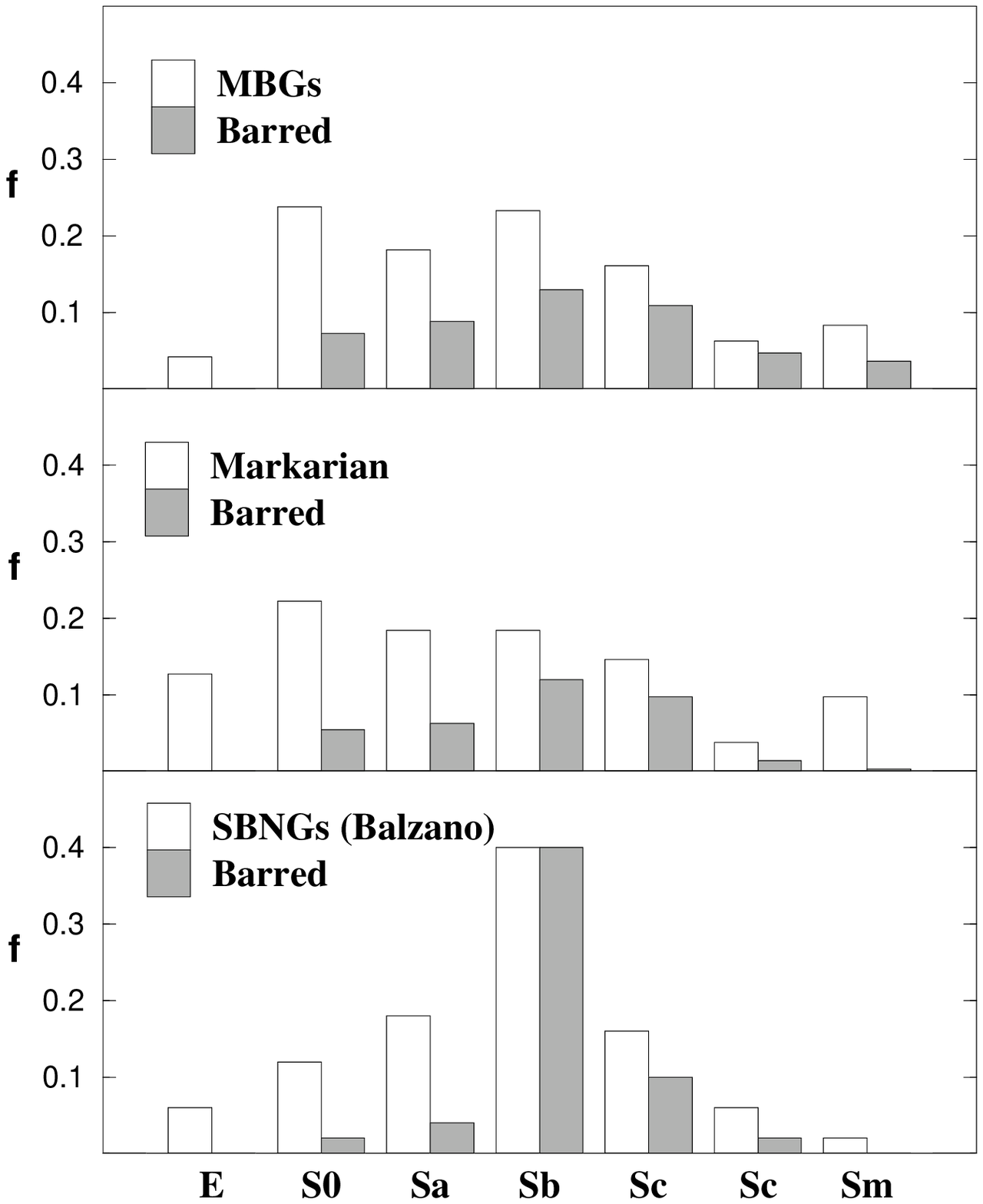]{The morphologies of the MBGs are compared to those of
the Markarian galaxies and the SBNGs from Balzano's sample (1983). The definition of
SBNGs, as defined by Balzano, should be enlarged to include a greater number of early--type spirals.} 

\clearpage

{\sc TABLE} 1. Third list of UV--bright galaxies.

{\sc TABLE} 2. Spectroscopic characteristics in the red.

{\sc TABLE} 3. Spectroscopic characteristics in the blue.

{\sc TABLE} 4. Metallicities and electron densities of MBGs.

{\sc TABLE} 5. CCD photometry of selected MBGs.

{\sc TABLE} 6. Correction for incompleteness.

\end{document}


\clearpage
\begin{deluxetable}{rcccclclc}
\tablecolumns{9}
\tablewidth{0pc}
\tablecaption{Third list of UV-bright galaxies}
\tablehead{
\colhead{Name}&\colhead{$\alpha$}&\colhead{$\delta$}&\colhead{B}&\colhead{$(U-B)$}&
\colhead{IRAS}&\colhead{$M_B$}&\colhead{Cross Identification}&\colhead{}\\
\colhead{MBG}&\colhead{(1950)}&\colhead{(1950)}&\colhead{(APM)}&\colhead{(APM)}&
\colhead{}&\colhead{(APM)}&\colhead{(NED)}&\colhead{}}
\startdata
00336--0347&00 33 36.1 &--03 47 48& 14.5&+0.2&no &\nodata&MCG--01--02--040 &\nodata\nl
00351--0433&00 35 06.7 &--04 33 29& 14.8&--0.5   &no &\nodata&\nodata&\nodata\nl
00405--7400&00 40 30.1 &--74 00 33& 14.6&--1.3   &yes&\nodata&\nodata&\nodata\nl
00407--0655&00 40 45.4 &--06 55 01& 15.4&--0.5   &yes&--18.9 &\nodata&\nodata\nl
00419--0317&00 41 58.2 &--03 17 26& 15.4&+0.4    &no &\nodata&\nodata&\nodata\nl
\tablevspace{0.2cm}
00456--0302&00 45 30.2 &--03 02 56& 11.7&+0.2&yes&--21.8&NGC 259&\nodata\nl
00463--0239&00 46 20.1 &--02 39 17& 14.7&--0.4&yes&--18.9&MRK 557&\nodata\nl
00476--0527&00 47 37.0 &--05 27 58& 12.6&--0.5&yes&--21.9&NGC 268&\nodata\nl
00485--0719&00 48 32.8 &--07 19 59& 13.2\tablenotemark{*}&\nodata &no&--18.6&NGC 275&n\nl
00496--4828&00 49 38.7 &--48 28 44& 15.3&--0.4&yes&--19.7&ESO 195--IG 010&\nodata\nl
\tablevspace{0.2cm}
00535--5044&00 53 30.1 &--50 44 05& 15.5&\phs0.0&yes&--18.5&ESO 195-- G 017&\nodata\nl
00550--0516&00 55 02.9 &--05 16 20& 14.8&--0.4&yes&--19.6&MCG--01--03--041  &\nodata\nl
00551--0513&00 55 10.1 &--05 13 09& 15.3&+0.5&yes&--18.9&MRK 966        &\nodata\nl
00556--0631&00 55 38.9 &--06 31 49& 15.0\tablenotemark{*}&\nodata&no &--19.5&MCG--00--03--050   &\nodata\nl
01025--0700&01 02 30.5 &--07 00 17& 14.9&--0.6&no &\nodata&MCG--01--03--086&\nodata\nl
\tablevspace{0.2cm}
01046--0416&01 04 37.7 &--04 16 34& 15.5&--0.1&yes&\nodata&\nodata&\nodata\nl
01047--0324&01 04 42.1 &--03 24 15& 15.1&--0.3&no &--19.3&MCG--01--03--092&\nodata\nl
01056--0449&01 05 39.5 &--04 49 17& 15.0&+0.6&yes&\nodata&\nodata&n\nl
01069--0307&01 06 53.8 &--03 07 11& 16.0&--0.6&no &--17.8&\nodata&\nodata\nl
01089--4743&01 08 57.9 &--47 43 32& 15.1&--0.5&yes&\nodata&\nodata&\nodata\nl
\tablevspace{0.2cm}
01117--6150&01 11 42.7 &--61 50 37& 15.1&+0.5&no &\nodata&\nodata&\nodata\nl
01137--5027&01 13 47.3 &--50 27 11& 15.4&+0.2&yes&--18.8&Fairall 294&n\nl
01179--6230&01 17 54.1 &--62 30 42& 15.4&--0.1&yes&\nodata&AM 0118--623&\nodata\nl
01196--6138&01 19 37.6 &--61 38 17& 15.0&+0.8&no &\nodata&ESO 113-- G 041&\nodata\nl
01196--6201&01 19 38.7 &--62 01 33& 14.4&\phs0.0&yes&\nodata&ESO 113--IG 043&\nodata\nl
\tablevspace{0.2cm}
01264--6141&01 26 26.0 &--61 41 48& 15.3&--0.8&no &\nodata&ESO 113-- G 049&\nodata\nl
01366--3010&01 36 41.5 &--30 10 42& 14.7&+0.3&no &--19.7&NGC 639&\nodata\nl
01400--3330&01 40 00.0 &--33 30 42& 15.0&--0.3&yes&--19.5&ESO 353-- G 033&\nodata\nl
01414--6243&01 41 29.9 &--62 43 24& 15.0&--0.0&yes&\nodata&ESO 080--IG 004&\nodata\nl
01457--6313&01 45 43.2 &--63 13 11& 15.6&--0.7&no &--15.9&ESO 080-- G 006&\nodata\nl
\tablevspace{0.2cm}
01578--6806&01 57 53.5 &--68 06 43& 13.5&--0.4&no &--18.0&NGC 802&n\nl
02043--5525&02 04 20.1 &--55 25 54& 15.5&+0.6&yes&--19.0&ESO 153--G 020&n\nl
02107--5403&02 10 42.8 &--54 03 45& 15.4&+0.2&no &--19.4&Fairall 378&\nodata\nl
02204--6450&02 20 24.6 &--64 50 14& 13.8&\phs0.0&yes&--20.7&ESO 081--G 008&\nodata\nl
02473--5008&02 47 22.6 &--50 08 44& 14.9&--0.2&yes&\nodata&ESO 199-- G 001&\nodata\nl
\tablevspace{0.2cm}
03012--5803&03 01 15.2 &--58 03 22& 14.6&+0.4&yes&--19.7&ESO 116--G 005&\nodata\nl
03019--2615&03 01 57.8 &--26 15 51&  8.3&+1.2&no&--23.4&NGC 1201&n\nl
03023--2739&03 02 23.6 &--27 39 15& 14.1&+0.2&yes&--20.6&IC 1876,&n\nl
03027--2742&03 02 46.6 &--27 42 24& 15.0\tablenotemark{*}&\nodata&yes&\nodata&ESO 417--IG 014&n\nl
03031--4907&03 03 09.1 &--49 07 27& 15.5&--1.1&yes&\nodata&ESO 199--IG 015&\nodata
\tablebreak
03037--2725&03 03 44.1 &--27 25 24& 15.4&+0.4&yes&\nodata&ESO 480-- G 030&\nodata\nl
03118--5732&03 11 48.7 &--57 32 38& 14.0&--0.7&yes&--16.9&ESO 116-- G 012&\nodata\nl
03142--2602&03 14 12.0 &--26 02 17& 14.9&+0.3&no &\nodata&MCG--04--08--055&\nodata\nl
03169--0618&03 16 57.8 &--06 18 08& 14.4&--0.2&yes&--18.0&MRK 1075       &\nodata\nl
03278--0424&03 27 49.4 &--04 24 48& 13.2&+0.2&yes&--22.0&KUG 0327-044   &\nodata\nl
\tablevspace{0.2cm}
03301--0455&03 30 07.7 &--04 55 09& 13.7&+0.3&yes&\nodata&KUG 0330-049   &\nodata\nl
03383--0359&03 38 19.1 &--03 59 24& 15.1&+0.2&yes&--18.6&MRK 1078       &\nodata\nl
03400--0427&03 40 03.1 &--04 27 27& 13.7&+0.7&yes&\nodata&IC 347         &\nodata\nl
03402--0632&03 40 12.6 &--06 32 24& 14.3&+0.4&yes&--20.4&MRK 1191       &\nodata\nl
03458--0646&03 45 57.7 &--06 46 45& 15.2&+0.3&yes&\nodata&MCG-01-10-038  &\nodata\nl
\tablevspace{0.2cm}
03505--4440&03 50 31.0 &--44 40 52& 13.6&--0.8&yes&--18.2&NGC 1476&\nodata\nl
03533--2817&03 53 18.6 &--28 17 01& 13.7\tablenotemark{*}&\nodata&yes&--17.8&IC 2007&\nodata\nl
03540--4229&03 54 01.1 &--42 29 49& 11.9\tablenotemark{*}&\phs0.0&yes&--18.3&NGC 1487&n\nl
03540--2636&03 54 04.3 &--26 36 58& 13.8&+0.2&\nodata&\nodata&\nodata&\nodata\nl
03590--4910&03 59 02.4 &--49 10 08& 13.6&--1.0&yes&\nodata&ESO 201-- G 014&\nodata\nl
\tablevspace{0.2cm}
04000--5250&04 00 00.7 &--52 50 49& 15.5&+0.7&yes&--20.1&IC 2028&\nodata\nl
04019--4332&04 01 54.1 &--43 32 11& 12.9&--0.7&yes&--17.7&NGC 1510&n\nl
04022--4329&04 02 15.6 &--43 29 06& 11.5&+0.4&yes&--18.9&NGC 1512&\nodata\nl
04026--3618&04 02 36.0 &--36 18 38& 15.1\tablenotemark{*}&\nodata&no&--16.3&ESO 359-- G 016&\nodata\nl
04030--4610&04 03 02.4 &--46 10 38& 15.0&\phs0.0&no&--16.2&ESO 250-- G 005&\nodata\nl
\tablevspace{0.2cm}
04048--5248&04 04 51.2 &--52 48 12& 13.6&--1.0&yes&--16.2&NGC 1552&\nodata\nl
04074--4538&04 07 28.3 &--45 38 52& 10.4&+0.4&yes&--21.0&IC 2035&\nodata\nl
04098--4608&04 09 49.6 &--46 08 05& 14.1&--0.8&yes&--20.0&AM 0409--460&\nodata\nl
04100--3258&04 10 03.9 &--32 58 42& 10.6&\nodata&no&--20.0&NGC 1531&\nodata\nl
04100--3301&04 10 05.6 &--33 01 24& 13.4&--1.5&yes&\nodata&\nodata&\nodata\nl
\tablevspace{0.2cm}
04101--3300&04 10 08.8 &--33 00 04& 11.2&+2.7&yes&--19.8&NGC 1532&\nodata\nl
04139--5104&04 13 58.1 &--51 04 12& 15.6&--0.9&yes&--17.9&ESO 201--IG 026&\nodata\nl
04163--5017&04 16 23.3 &--50 17 08& 11.3&--0.3&yes&--19.4&NGC 1556&n\nl
04189--5503&04 18 53.5 &--55 03 24& 10.6&+2.6&yes&--20.9&NGC 1566       &\nodata\nl
04210--4042&04 21 01.1 &--40 42 59& 14.2&+0.3&yes&--20.3&NGC 1572&\nodata\nl
\tablevspace{0.2cm}
04243--4323&04 24 23.4 &--43 23 05& 15.0&--0.2&yes&\nodata&\nodata&\nodata\nl
04253--5317&04 25 22.3 &--53 17 55& 14.5&--0.1&yes&--19.1&IC 2073&\nodata\nl
04259--4216&04 25 54.7 &--42 16 28& 13.0&--0.9&yes&--21.0&NGC 1585&\nodata\nl
04267--5510&04 26 45.3 &--55 10 07& 15.1&--1.0&no&\nodata&ESO-LV 1570321&\nodata\nl
04268--4321&04 26 52.9 &--43 21 02& 14.6&--0.5&yes&\nodata&\nodata&\nodata\nl
\tablevspace{0.2cm}
04287--5148&04 28 44.1 &--51 48 21& 15.0&+0.5&yes&--16.5&\nodata&\nodata\nl
04309--4947&04 30 55.9 &--49 46 52& 15.4&--0.5&yes&--16.6&ESO 202-- G 035&\nodata\nl
04337--4218&04 33 42.7 &--42 18 17& 15.0&--0.4&yes&\nodata&ESO 304-- G 002&\nodata\nl
04341--4035&04 34 11.3 &--40 35 38& 15.5&--0.2&yes&\nodata&\nodata&\nodata\nl
04350--4017&04 35 02.7 &--40 17 58& 15.5&--0.6&yes&\nodata&ESO 304--IG 005&n
\tablebreak
04425--4135&04 42 33.6 &--41 35 20& 14.4&\phs0.0&yes&\nodata&NGC 1660&\nodata\nl
04435--4140&04 43 30.0 &--41 40 09& 14.8&+0.9&no&\nodata&ESO 304-- G 019&\nodata\nl
20533--4410&20 53 23.0 &--44 10 44& 14.6&--0.2&yes&--19.5&NGC 6983       &\nodata\nl
21035--2436&21 03 30.9 &--24 36 48& 15.0&+0.4&yes&\nodata&NGC 7019&\nodata\nl
21065--4543&21 06 32.0 &--45 43 54& 14.9&+0.3&yes&\nodata&AM2106-454     &\nodata\nl
\tablevspace{0.2cm}
21068--3742&21 06 51.2 &--37 42 04& 17.0&+0.1&yes&\nodata&ESO 342--IG 013&n\nl
21097--3804&21 09 47.5 &--38 04 32& 15.4&--1.3&yes&--20.8&ESO 342-- G 022&\nodata\nl
21117--4724&21 11 45.3 &--47 24 44& 15.4&+0.5&yes&--18.7&NGC 7038       &\nodata\nl
21118--4701&21 11 49.9 &--47 01 51& 15.0&--0.2&no&--20.8&Fairall 961 &\nodata\nl
21142--2319&21 14 13.8 &--23 19 52& 15.0&+0.9&no&\nodata&ESO 530-- G 029&\nodata\nl
\tablevspace{0.2cm}
21169--2224&21 16 56.7 &--22 24 09& 14.5&\nodata&no&\nodata&ESO 599-- G 007&\nodata\nl
21194--3653&21 19 25.7 &--36 53 42& 14.5&+0.6&yes&--18.4&ESO 402-- G 026&\nodata\nl
21273--4439&21 27 21.0 &--44 39 00& 14.9&--0.2&yes&--20.9&ESO 287--IG 030&\nodata\nl
21317--5627&21 31 46.1 &--56 27 30& 15.2&+0.3&yes&--21.2&AM 2131--562&\nodata\nl
21331--5446&21 33 06.5 &--54 46 07& 11.3\tablenotemark{*}&\nodata&yes&--19.0&NGC 7090&\nodata\nl
\tablevspace{0.2cm}
21384--5300&21 38 25.8 &--53 00 06& 12.3&--1.8&yes&\nodata&IC 5125&\nodata\nl
21397--5255&21 39 42.8 &--52 55 10& 15.2&--0.3&yes&--19.0&ESO 188--IG 018&n\nl
21449--5512&21 44 56.8 &--55 12 13& 15.4&+0.2&no&--20.7&Fairall 591&\nodata\nl
21456--6056&21 45 37.0 &--60 56 45& 14.3&+0.2&yes&--18.8&NGC 7125&\nodata\nl
21488--5548&21 48 49.1 &--55 48 17& 14.7&--0.2&yes&--18.3&NGC 7140&\nodata\nl
\tablevspace{0.2cm}
21509--1331&21 50 59.4 &--13 31 53& 15.2&+0.2&no&\nodata&NPM16-13.054   &\nodata\nl
21515--6023&21 51 31.3 &--60 23 59& 15.5&--0.1&no&\nodata&\nodata&\nodata\nl
22051--5741&22 05 10.8 &--57 41 16& 13.5&+1.7&yes&--18.3&NGC 7205&\nodata\nl
22198--3227&22 19 50.6 &--32 27 08& 15.1&+1.0&no&--20.1&ESO 467-- G 046&\nodata\nl
22215--3356&22 21 33.4 &--33 56 24& 12.9\tablenotemark{*}&\nodata&yes&--20.3&NGC 7267&\nodata\nl
\tablevspace{0.2cm}
22220--3134&22 22 00.1 &--31 34 47& 14.8&+1.3&no&--20.4&ESO 467-- G 054&\nodata\nl
22228--3127&22 22 50.4 &--31 27 19& 15.0&+0.5&no&\nodata&NGC 7268&\nodata\nl
22233--3123&22 23 20.7 &--31 23 59& 13.8&+0.6&yes&--19.9&NGC 7277&\nodata\nl
22312--3239&22 31 12.0 &--32 39 18& 15.4&+0.5&yes&--18.0&ESO 405-- G 029&\nodata\nl
22333--3159&22 33 21.8 &--31 59 13& 15.0&+1.1&no&--20.1&ESO 468-- G 016&\nodata\nl
\tablevspace{0.2cm}
22467--4906&22 46 42.4 &--49 06 51& 15.1&+0.4&yes&--21.1&Fairall 358&\nodata\nl
22474--5142&22 47 26.6 &--51 42 47& 15.5&--0.2&yes&\nodata&\nodata&\nodata\nl
22514--3721&22 51 27.4 &--37 21 03& 15.1&--0.3&no&\nodata&ESO 406-- G 018&\nodata\nl
22522--3851&22 52 17.6 &--38 51 05& 15.2&\phs0.0&yes&--17.6&ESO 346-- G 014&\nodata\nl
22530--3427&22 53 05.2 &--34 27 35& 14.8&--0.4&yes&--20.5&ESO 406-- G 021&\nodata\nl
\tablevspace{0.2cm}
22530--3449&22 53 05.8 &--34 49 22& 13.7&--0.6&no&--22.9&ESO 406-- G 022&\nodata\nl
22538--3702&22 53 53.5 &--37 02 26& 14.7&--0.6&yes&--17.5&NGC 7418A&\nodata\nl
22543--3643&22 54 23.0 &--36 43 49&  8.6&+2.1&yes&--23.2&IC 1459&n\nl
22548--3421&22 54 48.8 &--34 21 21& 14.6&+0.3&no&\nodata&AM 2254--342&\nodata\nl
22550--3607&22 55 05.0 &--36 07 42& 13.0\tablenotemark{*}&\nodata&no&--18.7&IC 5270&\nodata
\tablebreak
22551--3747&22 55 08.8 &--37 47 06& 15.2&--0.8&no&\nodata&ESO 346--IG 020&n\nl
22565--4809&22 56 32.1 &--48 09 34& 14.4&--1.0&yes&\nodata&ESO 239--IG 006&n\nl
22566--3758&22 56 39.0 &--37 58 43& 15.4&--0.9&yes&--15.4&IC 5273&n\nl
22580--3538&22 58 01.8 &--35 38 23& 14.8&--0.8&yes&--17.1&IC 5269C&\nodata\nl
23004--3540&23 00 27.4 &--35 40 26& 15.2&--0.4&no&\nodata&\nodata&\nodata\nl
\tablevspace{0.2cm}
23164--6128&23 16 28.2 &--61 28 58& 14.5&\phs0.0&yes&\nodata&ESO 148-- G 004&\nodata\nl
23283--6112&23 28 20.9 &--61 12 30& 14.3&+0.2&yes&--18.9&ESO 148-- G 018&\nodata\nl
23289--6253&23 28 56.8 &--62 53 49& 15.5&+0.1&no&\nodata&\nodata&\nodata\nl
23335--3812&23 33 35.6 &--38 12 54& 14.4&+0.5&no&--15.4&NGC 7713&n\nl
23336--6333&23 33 37.3 &--63 33 57& 15.5&--0.1&yes&\nodata&\nodata&\nodata\nl
\tablevspace{0.2cm}
23361--5208&23 36 07.4 &--52 08 15& 14.7&--0.7&yes&--17.2&ESO 240-- G 012&\nodata\nl
23391--3654&23 39 11.1 &--36 54 11& 15.4&--0.4&no&\nodata&ESO 408--IG 017&n\nl
23417--3608&23 41 45.7 &--36 08 08& 15.5&+0.1&no&\nodata&\nodata&\nodata\nl
23473--6336&23 47 21.5 &--63 36 19& 14.5&--0.2&yes&\nodata&AM 2347--633&\nodata\nl
23493--6324&23 49 21.2 &--63 24 42& 15.3&--0.1&yes&\nodata&AM 2349--632&\nodata\nl
\enddata
\tablenotetext{*}{Galaxy not in the APM file, B magnitude taken from NED.}
\end{deluxetable}